\documentclass[prb,twocolumn,showpacs,preprintnumbers,amsmath,amssymb,superscriptaddress, bibnotes]{revtex4}

\usepackage{color}
\usepackage{graphicx}
\usepackage{dcolumn}
\usepackage{bm}

\begin{document}

\title{Suppression of superconductivity and structural phase transition against pressure in Ca$_{10}$(Ir$_{4}$As$_{8}$)(Fe$_{2-x}$Ir$_{x}$As$_{2}$)$_{5}$}

\author{Shunsaku~Kitagawa}
\email{shunsaku@science.okayama-u.ac.jp}
\author{Shingo~Araki}
\author{Tatsuo~C.~Kobayashi}
\affiliation{Department of Physics, Okayama University, Okayama 700-8530, Japan}
\affiliation{Research Center of New Functional Materials for Energy Production, Storage and Transport, 
Okayama University, Okayama 700-8530, Japan}
\author{Hiroyuki~Ishii}
\author{Kazunori~Fujimura}
\author{Daisuke~Mitsuoka}
\affiliation{Department of Physics, Okayama University, Okayama 700-8530, Japan}
\author{Kazutaka~Kudo}
\author{Minoru~Nohara}
\affiliation{Department of Physics, Okayama University, Okayama 700-8530, Japan}
\affiliation{Research Center of New Functional Materials for Energy Production, Storage and Transport, 
Okayama University, Okayama 700-8530, Japan}

\date{\today}

\newcommand{\red}[1]{\textcolor{red}{#1}}

\begin{abstract}
We measured the pressure dependence of in-plane resistivity $\rho_{ab}$ in the recently-discovered iron-based superconductor Ca$_{10}$(Ir$_{4}$As$_{8}$)(Fe$_{2-x}$Ir$_{x}$As$_{2}$)$_{5}$, which shows a unique structural phase transition in the absence of magnetic ordering, with a superconducting transition temperature $T_{\rm c}$ = 16~K and structural phase transition temperature $T_{\rm s}$ $\simeq$ 100~K at ambient pressure.
$T_{\rm c}$ and $T_{\rm s}$ are suppressed on applying pressure and disappear at approximately 0.5~GPa, suggesting a relationship between superconductivity and structure.
Ca$_{10}$(Ir$_{4}$As$_{8}$)(Fe$_{2-x}$Ir$_{x}$As$_{2}$)$_{5}$ is a rather rare example in which the superconductivity appears only in a low-temperature ordered phase.
The fact that the change in the crystal structure is directly linked with superconductivity suggests that the crystal structure as well as magnetism are important factors governing superconductivity in iron pnictides.
\end{abstract}


\abovecaptionskip=-5pt
\belowcaptionskip=-10pt

\maketitle

\section{Introduction} 
Superconductivity often appears with the suppression of a low-temperature phase such as magnetic and charge ordered phases\cite{T.Yamauchi_PRL_2002,E.Morosan_Nature_2006,K.Kanoda_JPSJ_2006,P.A.Lee_RMP_2006,B.Sipos_NatMater_2008,C.Pfleiderer_RMP_2009,D.Aoki_JPSJ_2014,S.Pyon_JPSJ_2012,T.Yajima_JPSJ_2013,S.Kitagawa_JPSJ_2013,H.Kotegawa_JPSJ_2014}.
Such superconductors are considered to be associated with the instability of the order parameter.
Iron pnictides exhibit a variety of structural phase transitions (SPTs), which frequently couple with magnetism, and hence superconductivity.

\begin{figure}[!t]
\vspace*{10pt}
\begin{center}
\includegraphics[width=7.5cm,clip]{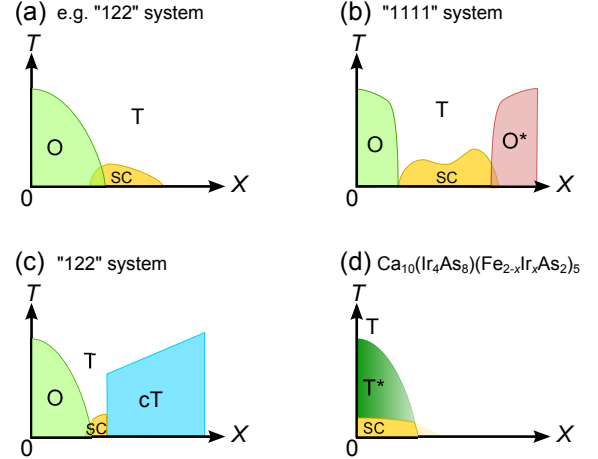}
\end{center}
\caption{(Color online) Schematic images of the phase diagrams in several iron pnictides;
$X$ represents tuning parameters such as pressure $p$ and chemical doping $x$.
T, T$^*$, cT, O, and O$^{*}$ indicate tetragonal, low temperature phase in Ca$_{10}$(Ir$_{4}$As$_{8}$)(Fe$_{2-x}$Ir$_{x}$As$_{2}$)$_{5}$, collapsed tetragonal, orthorhombic, and orthorhombic phase in heavily H-doped region, respectively.
}
\label{Fig.1}
\end{figure}

There are three typical examples of SPTs in iron pnictides, as illustrated in Fig.\ref{Fig.1}:
(a) a tetragonal (T) to orthorhombic (O) phase transition is observed in ``1111'' and ``122'' systems\cite{Y.Kamihara_JACS_2008,K.Ishida_JPSJ_2009,J.Paglione_Naturephys_2010}.
This SPT enhances in-plane antiferromagnetic (AFM) fluctuation\cite{Y.Nakai_PRB_2012} and is followed by the stripe-type AFM transition.
Such a transition can be suppressed by pressure or chemical substitution, and superconductivity appears when this type of transition disappears.
Therefore, such superconductivity has been understood using the magnetic quantum critical point and/or structural quantum critical point senario\cite{I.I.Mazin_PRL_2008,K.Kuroki_PRL_2008,V.Cvetkovic_EPL_2009,H.Kontani_PRL_2010}.
(b) A tetragonal (T) to orthorhombic (O$^*$) phase transition is observed in heavily H-doped ``1111'' systems.
Recently, it was discovered that heavily electron-doped regions in the ``1111'' system can be accessed by substituting H for O, and that the superconducting (SC) transition temperature $T_{\rm c}$ in the ``1111'' system has a two-dome structure as a function of electron doping\cite{S.Iimura_NC_2012}.
Moreover, an SPT is discovered in electron-overdoped regions\cite{M.Hiraishi_NatPhys_2014}.
Below the SPT temperature $T_{\rm s}$, the crystal has an orthorhombic $Aem2$ structure without inversion symmetry, in contrast to the $Cmme$ structure with inversion symmetry for case (a).
This SPT is also accompanied by AFM magnetic order.
The magnetic structure is another stripe-type AFM state, and the ordered moment is larger than that in case (a).
This AFM phase in heavily-doped regions might be related to the wide SC region in the ``1111'' system\cite{M.Hiraishi_NatPhys_2014}.
(c) A tetragonal (T) or orthorhombic (O) to collapsed tetragonal (cT) phase transition is observed in ``122'' system.
In ``122'' system, another SPT to the cT phase is observed on applying pressure or with chemical substitution\cite{M.S.Torikachvili_PRL_2008,A.I.Goldman_PRB_2009,S.Kasahara_PRB_2011,M.Danura_JPSJ_2011,R.Mittal_PRB_2011,W.O.Uhoya_JPCM_2011}.
After this SPT, the lattice parameter $c$ is significantly reduced, and the Fermi surface becomes three dimensional.
Consequently, Fermi liquid transport and loss of the magnetic moment and superconductivity are observed in the collapsed tetragonal phase, in contrast to non-Fermi liquid transport and spin-fluctuation-mediated superconductivity in the uncollapsed tetragonal phase.
The common features among the above three SPTs are the strong coupling between the structure and magnetism, and the disappearance of superconductivity in the low-temperature phase.
The superconductivity primarily appears in the tetragonal phase associated with AFM correlations in iron pnictides.

Very recently, a novel type of SPT was reported in Ca$_{10}$(Ir$_{4}$As$_{8}$)(Fe$_{2-x}$Ir$_{x}$As$_{2}$)$_{5}$\cite{K.Kudo_SR_2013} [case (d) in Fig.\ref{Fig.1}].
A synchrotron X-ray diffraction study revealed that the SPT, which corresponds to the doubling of the lattice parameter $c$, occurs at approximately 100~K, and can be attributed to a shift of half of the Ir ions\cite{N.Katayama_JPSJ_2014}.
Note that this SPT is not related to the Fe moment and that no magnetic order is observed.
After the SPT, SC transition is observed at $T_{\rm c}$ = 16~K.
Ca$_{10}$(Ir$_{4}$As$_{8}$)(Fe$_{2-x}$Ir$_{x}$As$_{2}$)$_{5}$ is a suitable material to investigate the relationship between structure and superconductivity without considering magnetism.
Applying pressure is suitable for suppressing the SPT because the crystal volume in the low-temperature phase (LTP) is larger than that extrapolated from the value in the high-temperature phase (HTP).

In this study, we measured the temperature ($T$) dependences of in-plane resistivity $\rho_{ab}$ under pressure up to $\sim$ 4.2~GPa in single-crystalline Ca$_{10}$(Ir$_{4}$As$_{8}$)(Fe$_{2-x}$Ir$_{x}$As$_{2}$)$_{5}$.
The superconducitivity is suppressed by pressure and disappears at approximately 0.5~GPa.
The SPT also disappears at this pressure, suggesting a relationship between the superconductivity and structure.
Ca$_{10}$(Ir$_{4}$As$_{8}$)(Fe$_{2-x}$Ir$_{x}$As$_{2}$)$_{5}$ is the only example among iron-based superconductors in which the superconductivity appears only in the LTP.
Thus, Ca$_{10}$(Ir$_{4}$As$_{8}$)(Fe$_{2-x}$Ir$_{x}$As$_{2}$)$_{5}$ is important for understanding the role of the structure in iron-based superconductors.

\section{Experimental}

Single-crystalline Ca$_{10}$(Ir$_{4}$As$_{8}$)(Fe$_{2-x}$Ir$_{x}$As$_{2}$)$_{5}$ samples were grown by heating a mixture of Ca, FeAs, IrAs$_2$, and Ir powders in a ratio of Ca:Fe:Ir:As = 10:10:4:18 or 10:26:14:40. 
The mixture was placed in an alumina crucible and sealed in an evacuated quartz tube. 
Manipulation was performed in a glove box filled with argon gas. 
The ampules were heated at 700$^{\rm o}$C for 3 h and then at 1100-1150$^{\rm o}$C for 10-40~h, after which they were quenched in cold water\cite{K.Kudo_SR_2013}.
The sample is identical to that in a previous report\cite{K.Kudo_SR_2013}.
The $\rho_{ab}$ measurements at high pressures of up to $\sim$ 4.2 GPa were performed using an indenter cell\cite{T.C.Kobayashi_RSI_2007}.
The electrical resistivity was measured using a standard four-probe method. 
We used Daphne 7474 as a pressure-transmitting medium\cite{K.Murata_RSI_2008}. 
The applied pressure was estimated from the $T_{\rm c}$ of the lead manometer.

\begin{figure}[!b]
\vspace*{-0pt}
\begin{center}
\includegraphics[width=7.5cm,clip]{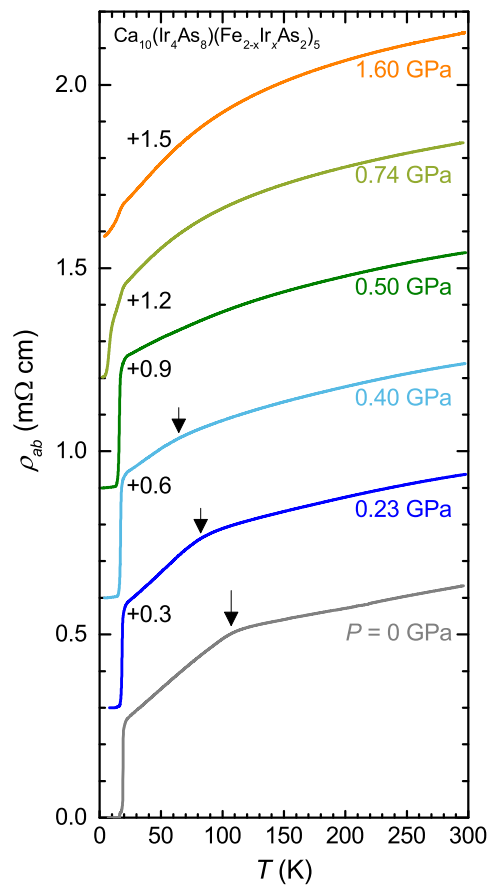}
\end{center}
\caption{(Color online) $T$ dependence of in-plane resistivity $\rho_{ab}$ under different pressures in Ca$_{10}$(Ir$_{4}$As$_{8}$)(Fe$_{2-x}$Ir$_{x}$As$_{2}$)$_{5}$.
The arrows indicate the structural phase transition temperature $T_{\rm s}$ defined by a  negative peak of $\partial^2 \rho_{ab}/\partial T^2$.
}
\label{Fig.2}
\end{figure}


\section{Results and Discussion}

\begin{figure}[!tb]
\vspace*{-0pt}
\begin{center}
\includegraphics[width=8.5cm,clip]{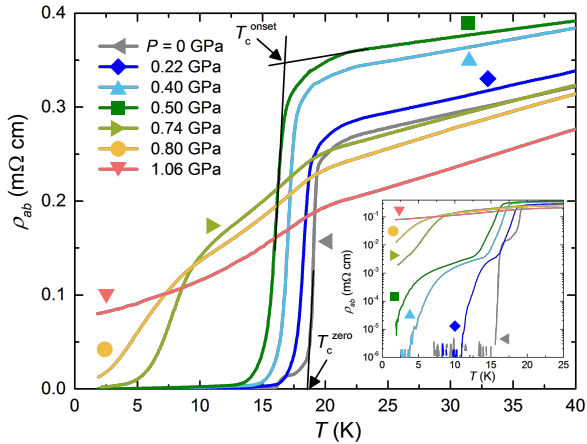}
\end{center}
\caption{(Color online) $T$ dependence of $\rho_{ab}$ below 40~K. 
$T_{\rm c}^{\rm zero}$ and $T_{\rm c}^{\rm onset}$ are defined as shown in the figure.
$T_{\rm c}$ is suppressed on increasing the pressure and zero resistivity is not observed down to 2~K at above 0.5~GPa.
The inset shows $\log\rho$ versus $T$.}
\label{Fig.3}
\end{figure}

\begin{figure}[!tb]
\vspace*{-0pt}
\begin{center}
\includegraphics[width=8cm,clip]{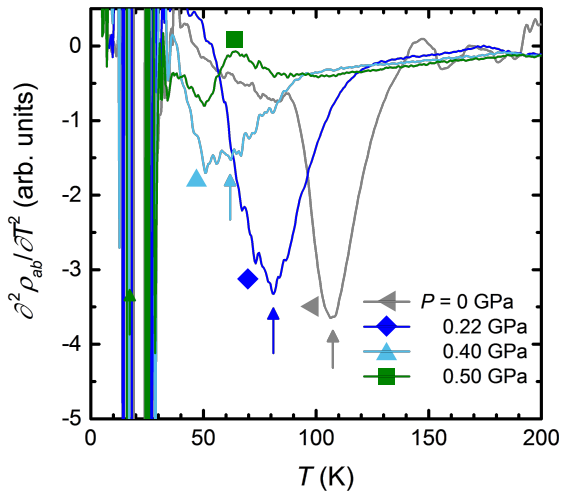}
\end{center}
\caption{(Color online) $T$ dependence of $\partial^2 \rho_{ab}/\partial T^2$.
A negative peak of $\partial^2 \rho_{ab}/\partial T^2$ corresponds to a kink in the resistivity.
Thus, $T_{\rm s}$ is defined as this negative peak of $\partial^2 \rho_{ab}/\partial T^2$.
$T_{\rm s}$ decreases with increasing pressure and disappears above 0.5~GPa.
}
\label{Fig.4}
\end{figure}

\begin{figure}[!tb]
\vspace*{0pt}
\begin{center}
\includegraphics[width=8.5cm,clip]{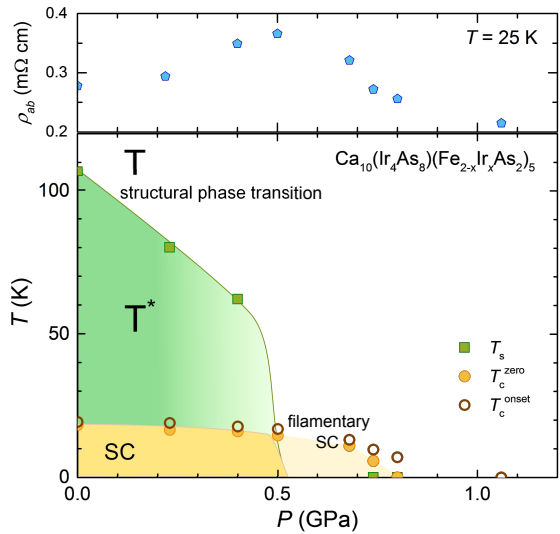}
\end{center}
\caption{(Color online)(Upper panel) $P$ dependence of $\rho_{ab}$ at 25~K.
A peak at 0.5~GPa suggests the suppression of the SPT at $\sim$ 0.5~GPa.
(Lower panel) $P$-$T$ phase diagram for Ca$_{10}$(Ir$_{4}$As$_{8}$)(Fe$_{2-x}$Ir$_{x}$As$_{2}$)$_{5}$.
Circles represent $T_{\rm c}$ and squares represent $T_{\rm s}$.
$T_{\rm c}^{\rm zero}$ and $T_{\rm c}^{\rm onset}$ are defined as shown in Fig.~\ref{Fig.3}.
$T_{\rm s}$ is defined as a negative peak of $\partial^2 \rho_{ab}/\partial T^2$.
$T_{\rm c}$ and $T_{\rm s}$ are suppressed on applying pressure and disappear at $\sim$ 0.5~GPa.
}
\label{Fig.5}
\end{figure}

Figure~\ref{Fig.2} shows the $T$ dependence of $\rho_{ab}$ in  Ca$_{10}$(Ir$_{4}$As$_{8}$)(Fe$_{2-x}$Ir$_{x}$As$_{2}$)$_{5}$ under different pressures.
At ambient pressure, $\rho_{ab}$ shows a kink at $\sim$ 100~K, which corresponds to the SPT.
On cooling, $\rho_{ab}$ becomes zero at 16~K.
$T_{\rm c}^{\rm zero}$ and $T_{\rm c}^{\rm onset}$ are defined as shown in Fig.~\ref{Fig.3}.
$T_{\rm c}$ is suppressed on increasing the pressure, and zero resistivity is not observed down to 2~K at above 0.5~GPa as shown in the inset of Fig.~\ref{Fig.3}, although the anomaly corresponding to the SC transition is observed above 0.5~GPa.
It seems that a part of the sample remains in the LTP, probably because of a first-order SPT, and the sample in the LTP region shows superconductivity.
The SPT in Ca$_{10}$(Ir$_{4}$As$_{8}$)(Fe$_{2-x}$Ir$_{x}$As$_{2}$)$_{5}$ should be first-order since the variation in crystal symmetry does not occur via the SPT.
On increasing the pressure further, the SC transition does not reappear up to 4.2~GPa.
A small kink is observed at approximately 20~K even above 1~GPa, which might correspond to the SC transition of a part of sample due to the pressure distribution.
Next, we focus on the SPT.
To infer $T_{\rm s}$, the $T$ dependence of $\partial^2 \rho_{ab}/\partial T^2$ is shown in Fig.~\ref{Fig.4}.
A negative peak of $\partial^2 \rho_{ab}/\partial T^2$ corresponds to a kink in the resistivity, and thus, we define $T_{\rm s}$ as this negative peak of $\partial^2 \rho_{ab}/\partial T^2$.
$T_{\rm s}$ decreases with increasing pressure.
At 0.5~GPa, tiny anomaly instead of a clear peak is observed in $\partial^2 \rho_{ab}/\partial T^2$ and we can not define $T_{\rm s}$, suggesting the disappearance of the SPT at around 0.5~GPa.
In addition, $P$ dependence of $\rho_{ab}$ at 25~K shows a peak at 0.5~GPa as shown in the upper panel of Fig.~\ref{Fig.5}, which also supports the suppression of the SPT at $\sim$ 0.5~GPa.

The $P$-$T$ phase diagram for Ca$_{10}$(Ir$_{4}$As$_{8}$)(Fe$_{2-x}$Ir$_{x}$As$_{2}$)$_{5}$ is shown in Fig.~\ref{Fig.5}.
$T_{\rm c}$ and $T_{\rm s}$ are suppressed on applying pressure and disappear at $\sim$ 0.5~GPa, suggesting a coupling between the superconductivity and structure.
Note that resistivity measurements are sensitive to superconductivity and insensitive to the SPT.
Therefore, the $P$ dependence of X-ray diffraction and specific heat measurements are strongly desired for precise determination of $P$-$T$ phase diagram.
The appearance of superconductivity in only the low-temperature phase is quite rare, even in strongly-correlated electron systems.
To the best of our knowledge, UGe$_2$ and UIr, in which superconductivity appears only in the ferromagnetic state, are the only example for such a system\cite{S.S.Saxena_Nature_2000,T.C.Kobayashi_JPSJ_2007}.
In comparison with other iron-based superconductors\cite{J.Paglione_Naturephys_2010,G.R.Stewart_RMP_2011}, the superconductivity in Ca$_{10}$(Ir$_{4}$As$_{8}$)(Fe$_{2-x}$Ir$_{x}$As$_{2}$)$_{5}$ is considerably sensitive against pressure.
For example, in BaFe$_2$As$_2$, superconductivity appears in the pressure range between 4 and 7~GPa\cite{K.Matsubayashi_JPSJ_2009}, and superconductivity exists up to 20~GPa in the ``1111'' system\cite{H.Takahashi_JPSJSC_2008,H.Takahashi_Nature_2008}.
Moreover, the initial slope of $T_{\rm c}$ in isostructural Ca$_{10}$(Pt$_{4}$As$_{8}$)(Fe$_{2-x}$Pt$_{x}$As$_{2}$)$_{5}$ with $T_{\rm c}$ = 38~K is $\sim$~-~0.9~K/GPa\cite{M.Nohara_SSC_2012}.
In addition to the rapid suppression of $T_{\rm c}$, the $P$-$T$ phase diagram of Ca$_{10}$(Ir$_{4}$As$_{8}$)(Fe$_{2-x}$Ir$_{x}$As$_{2}$)$_{5}$ is considerably different from that of other iron pnictides.
In particular, both the LTP and HTP are paramagnetic in Ca$_{10}$(Ir$_{4}$As$_{8}$)(Fe$_{2-x}$Ir$_{x}$As$_{2}$)$_{5}$, implying that superconductivity is independent of magnetism, whereas magnetism plays an important role in superconductivity in other iron pnictides.
Therefore, a detailed study of Ca$_{10}$(Ir$_{4}$As$_{8}$)(Fe$_{2-x}$Ir$_{x}$As$_{2}$)$_{5}$ gives unique information regarding the role of structure in iron pnictides.

Finally, we discuss the effect of the structure on superconductivity in Ca$_{10}$(Ir$_{4}$As$_{8}$)(Fe$_{2-x}$Ir$_{x}$As$_{2}$)$_{5}$.
Superconductivity does not appear in the HTP, although $T_{\rm c}$ = 38~K is observed in the isostructural Ca$_{10}$(Pt$_{4}$As$_{8}$)(Fe$_{2-x}$Pt$_{x}$As$_{2}$)$_{5}$ without an SPT\cite{S.Kakiya_JPSJ_2011,N.Ni_PNAS_2011,C.Lohnert_ACIE_2011,M.Nohara_SSC_2012}.
A remarkable difference between the Pt-system and the Ir-system is the contribution of the 5$d$ electrons in the spacer layers at the Fermi energy ($E_{\rm F}$).
The band-structure calculations for both the Pt-system and the Ir-system predict a finite contribution of the 5$d$ electrons in the spacer layers for the density of states (DOS) at $E_{\rm F}$, in contrast to the negligible partial DOS at $E_{\rm F}$ of the spacer layers for other iron-based superconductors\cite{C.Lohnert_ACIE_2011,K.Kudo_SR_2013}.
In the Pt-system, however, angle-resolved photoemission spectroscopy (ARPES) measurements show that the contribution of the Pt 5$d$ states at $E_{\rm F}$ is rather small even though it may exist as predicted by the theory, indicating that a good two dimensional (2D) Fermi surface is formed by Fe 3$d$ electrons\cite{X.P.Shen_PRB_2013}.
On the other hand, the ARPES measurements indicate a large contribution of Ir 5$d$ electrons to the DOS at $E_{\rm F}$ in Ca$_{10}$(Ir$_{4}$As$_{8}$)(Fe$_{2-x}$Ir$_{x}$As$_{2}$)$_{5}$, which is suggestive of the three-dimensional Fermi surface in the HTP\cite{K.Sawada_PRB_2014}.
Moreover, the ARPES measurements indicate that Ir 5$d$ electronic states in the LTP become glassy probably because of atomic disorder related to the SPT.
Thus, the interlayer coupling becomes weak and, consequently, the Fermi surface in the LTP becomes 2D, as commonly seen in various iron pnictides\cite{K.Sawada_PRB_2014}.
This might be why superconductivity only appears in the LTP.
It seems that the 2D Fermi surface is important for high-$T_{\rm c}$ superconductivity.
Band structure calculations for the LTP might be helpful to discuss the role of Fermi surface for superconductivity more precisely.

\section{Summary}

In conclusion, we measured the in-plane resistivity $\rho_{ab}$ under pressure in the recently-discovered iron-based superconductor Ca$_{10}$(Ir$_{4}$As$_{8}$)(Fe$_{2-x}$Ir$_{x}$As$_{2}$)$_{5}$ with $T_{\rm c}$ = 16~K and $T_{\rm s}$ $\simeq$ 100~K.
The SC transition and SPT are suppressed on applying pressure and disappear at approximately 0.5~GPa, suggesting a connection between  the superconductivity and structure.
Since this SPT is not related to the magnetism of the Fe 3$d$ moment, the crystal structure might play crucial role in the SC mechanism in iron pnictides.
The comparison between the Pt-system and the Ir-system suggests the importance of the 2D Fermi surface in high-$T_{\rm c}$ superconductivity.
Ca$_{10}$(Ir$_{4}$As$_{8}$)(Fe$_{2-x}$Ir$_{x}$As$_{2}$)$_{5}$ provides a new avenue to investigate the relationship between the superconductivity and structure in iron pnictides.
A further study about the variation in the structure and the electronic state as the function of pressure is strongly required.


\section*{Acknowledgments}
This work was partially supported by Okayama University cryogenic center, Grants-in-Aid for Scientific Research (B) (26287082) and (C) (25400372) 
from the Japan Society for the Promotion of Science (JSPS) and the Funding Program for World-Leading Innovative R\&D on Science and Technology (FIRST Program) from JSPS. 


%

\end{document}